\documentclass[%
 reprint,
showpacs,
showkeys,
 amsmath,amssymb,
 aps,
 pra,
floatfix,
]{revtex4-1}

\usepackage{graphicx}
\usepackage{dcolumn}
\usepackage{bm}
\usepackage{epstopdf}
\usepackage{color,soul}

\usepackage[mathlines]{lineno}

\begin{document}

\preprint{APS/123-QED}

\title{Quantum simulation of ZnO nanowire piezotronics} 

\author{Leisheng Jin}
 
\author{Lijie Li}%
 \email{l.li@swansea.ac.uk}
\affiliation{%
 College of Engineering, Swansea University, Swansea, SA2 8PP, UK\\
\\ 
}%

\date{June 4 2015}

\begin{abstract}
We address the problem of quantum transport in a nanometre sized two-terminal ZnO device subject to an external strain. The two junctions formed between the electrodes and the ZnO are generally taken as Ohmic and Schottky type, respectively. Unlike the conventional treatment to the piezopotential, we treat it as a potential barrier which is only induced at the interfaces. By calculating the transmission coefficient of a Fermi-energized electron that flows from one end to the other, it is found that the piezopotential has the effect of modulating the voltage threshold of the current flowing. The calculations are based on the quantum scattering theory. The work is believed to pave the way for investigating the quantum piezotronics. 

\end{abstract}

                              
\maketitle

\section{Introduction }
Piezotronics, coined by Zhonglin Wang in 2007 \cite{ref1}, refers to the fabricated electronics whose charge transport behaviour across a metal/semiconductor interface or a p-n junction can be tuned by inner-crystal piezopotential that plays the role of the gate voltage. ZnO as a piezoelectric material possessing two important properties, i.e., semiconductive and piezoelectric, has been extensively studied as one of the most ideal candidates for piezotronic devices, and many applications materialized by ZnO such as actuators, sensors and energy harvesting devices have been achieved. For example, pH, glucose and protein sensors fabricated using metal-ZnO micro/nanowire-metal (MSM) structure were realized in \cite{ref2} \cite{ref3} \cite{ref4}. Instead of using a single nanowire, wearable devices and pressure mapping sensors fabricated by ZnO nanowire arrays were explored in \cite{ref5} \cite{ref6} \cite{ref7} as well. Resistive switching devices were also achieved using ZnO nanowire arrays \cite{refli}. 

The fundamental physics behind the nanogenerator based on the ZnO nanowire has previously been discussed. The Schottky theory to explain the tunability of current transport in the piezotronics is deemed as the mainstream. It was argued that the piezopotential induced at the interface of the M$-$S is behaving as a potential modulator (gate voltage) which exists in a certain width, and the framework has been systematically presented by Zhang et al. in \cite{ref8}, and the theory was further developed into a new branch called piezo-phototronics by Liu et al. in \cite{ref9}. These two studies have provided reasonable explanations for the experiments. However, the theory constructed by the above literatures is still under the consideration of classical physics, i.e., the classical current-voltage formula in the semiconductor physics for calculating the current have been employed  \cite{ref8}. This is satisfactory for investigating the device which has a relatively larger scale \cite{ref10} (L $\gg$ de Broglie wavelength of an electron) but may not be very appropriate for the scale when the quantum effect cannot be ignored (for example: L $\sim$ $\lambda$) . In the latter case, the quantum effect will affect the working status of the device, and some unexpected quantum phenomenons emerge. Therefore, constructing the theory of the electron transport in piezotronics with the contents of quantum physics is becoming desirable. Especially nowadays, the devices are drastically decreased in size with the rapid development of fabrication techniques. The development of the theory will benefit to the exploration of the quantum piezotronics. In this work, investigation has been conducted to understand the quantum physics of the electron transport in piezotronic nano devices. By taking M-S-M ZnO structure that has ballistic length as a paradigm, we postulate the theoretical work for calculating the electrical current in the two-terminal device. We treat the electron transport from the perspective of quantum scattering theory. Instead of seeing the electron as a classical partial, the electron is treated as a quantum wave. The piezopotential induced at the interface were assumed to be taking up a certain width \cite{ref8}, while in this work the piezopotential is calculated only at the interface, which is believed to be more accurate, in particular, for the very short device. The work is believed to shed some light on understanding the next-generation quantum piezotronic devices. The theoretical analysis will be given in the section 2, where the basic background of the model and the fundamental view of the electron transport in the two-terminal device are described. Numerical simulation that is based on the reasonable parameters will be conducted in the section 3, and finally the conclusion is presented in the section 4.

\section{Theoretical Analysis}
\begin{figure}
\includegraphics[width=5cm]{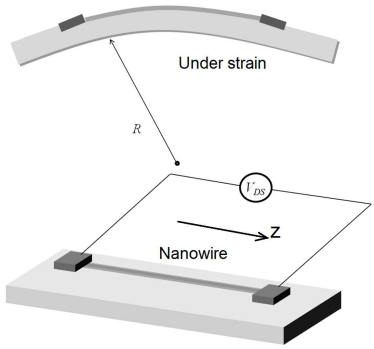}
\caption{\label{fig1} Schematic of the ZnO device.}
\end{figure}

\begin{figure}[b]
\includegraphics[width=5cm]{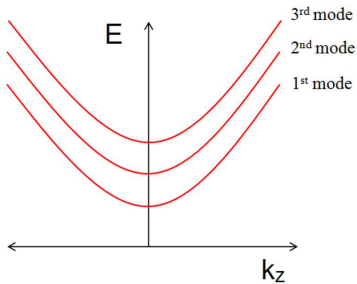}
\caption{\label{fig2} Energy modes of electrons in ZnO wire.}
\end{figure}

In the device (schematically shown in Fig.~\ref{fig1}), the ZnO nanowire is seen as a quantum wire, which is very short in $z$-direction (about 80nm) and has an extremely small area of cross section in $x-y$ plane. The electron transport in $z$-direction can be seen as ballistic, and only a few electron energy states exist in $x-y$ plane. For an electron in quantum wire, its energy is given by:

\begin{align}
&E=\frac{\pi^2 \hbar^2}{2m}\left( \frac{n_{x}^2}{L_{x}^2}+\frac{n_{y}^2}{L_{y}^2}+\frac{\hbar^2 k_{z}^2}{2m} \right) \nonumber\\
&n_{x},n_{y}=1,2,...
\label{eq:1}
\end{align}\noindent
where the electron in $z$-direction is modelled by plane waves with wave number $k_{z}$. $\hbar$ is the Planck's constant divided by $\pi$. $n_{x}$ and $n_{y}$ are energy quantum number in $x$ and $y$ directions, respectively. $m$ is the effective mass of the electron. $L_{x}$ and $L_{y}$ are the lengths in $x-y$ plane. Three modes of the energy in quantum wire are plotted in Fig.~\ref{fig2}. The electrons in the nanowire will occupy the energy level from low to high, and in the equilibrium state the highest occupied energy is called quasi Fermi levels, represented by $F+$ for electrons with $+k$ and $F-$ for the electrons with $-k$.

\begin{figure}
\includegraphics[width=8cm]{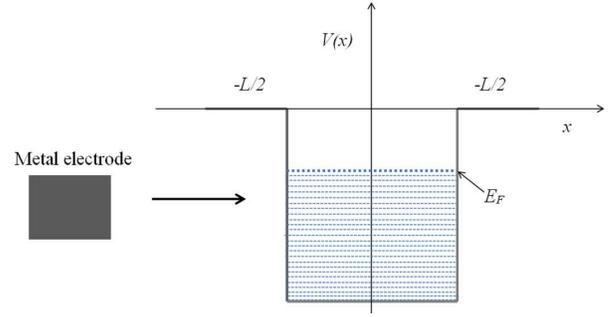}
\caption{\label{fig3} Energy distribution of electrons in the electrodes.}
\end{figure}

\begin{figure}
\includegraphics[width=7cm]{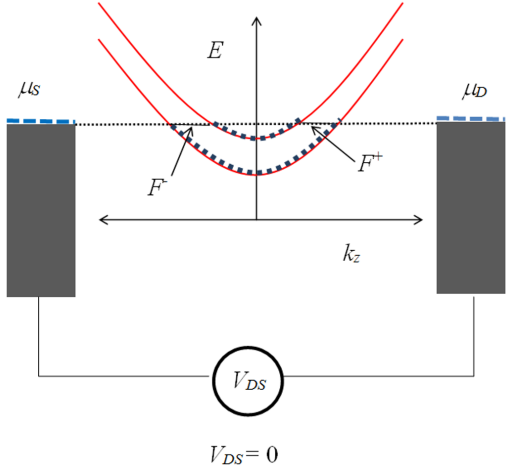}
\caption{\label{fig4} Electron states when the device is in equilibrium.}
\end{figure}

\begin{figure}
\includegraphics[width=7cm]{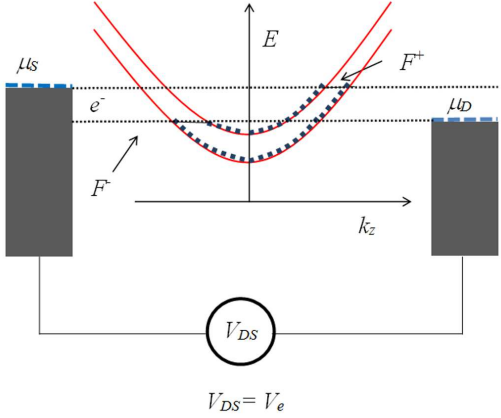}
\caption{\label{fig5} Electron states when the device is applied a voltage.}
\end{figure}

The left (Source) and right (Drain) electrodes of the device are assumed to be reservoirs of electrons, i.e., the electrons are kept presumably in equilibrium state, even under an applied voltage. The potential inside of the electrodes is approximately constant while the potential at the boundaries has a step which confines the electrons. The whole potential profile of the electrodes can be seen as a finite square well as shown in Fig.~\ref{fig3}. The allowed kinetic energy and wave functions of electrons inside of the electrodes can be obtained straightforwardly by solving the Schrodinger equation, as:

\begin{align}
&T_{n}=\frac{\hbar^2 k_{n}^2}{2m}=\frac{n^2 \pi^2 \hbar^2}{2mL^2} \nonumber\\
&\psi_{n}(z)=\sqrt{2/L} \sin k_{n} z=\frac{1}{i\sqrt{2L}}(e^{ik_{n}z}-e^{-ik_{n}z})
\label{eq:2}
\end{align}\noindent
where $L$ is the length of the electrodes in $z$-direction and $n$ ($n$=1,2,3…) is the energy quantum number. $k_{n}$ is $z$-direction component of wave vector at energy state $n$. The electrons occupy the energy state according to the Pauli exclusion principle that demands one energy state can only be taken by two electrons if considering the spin. The highest energy state occupied by electrons is Fermi energy $E_{F}$ or so-called chemical potential $\mu$. When the electrodes and the ZnO nanowire are connected and there is no bias applied to $S$ and $D$, the Fermi energy must be constant through the device. Otherwise, there will be current flowing. This scenario is plotted in Fig.~\ref{fig4}, where the chemical potential $\mu_{S}$, $F+$, $F-$ and $\mu_{D}$ are being flat. The number of electrons with $k+$ wavenumber equals to the number of the electrons with $k-$ wavenumber so there is not current. If a DC voltage $V_{e}$ is applied to the source ($S$) and drain ($D$), the electron equilibrium state will be broken. As shown in Fig.~\ref{fig5} the unbalance between the electrons with $k+$ and $k-$ occurs with $F+$ is higher than $F-$, which leads to some uncompensated electrons that have energy between $F+$ and $F$- flowing into the drain, i.e., current induced. The quasi Fermi level now has the following relations with chemical potential, which are: $F+=\mu_{S}$  and $F-=\mu_{D}$. Also, the relation of the shift between the potential energies of electrodes and the applied $V_{e}$ is given by $\mu_{D}-\mu_{S}=-qV_{e}$ , where $q$ is the charge of a single electron. 

Under an applied voltage $V_{e}$, the current in a ballistic quantum wire is given by the famous Landauer formula, which is:

\begin{align}
I_{in}=-q^2 \frac{V_{e}}{\pi \hbar}
\label{eq:3}
\end{align}\noindent
$I_{in}$ is the current brought by the uncompensated electrons. However, in this device, the quantum wire is made by ZnO, which is semiconducting material, and the electrodes are made of metals such as Ag. When a two-terminal device is fabricated, the M-S interfaces would be formed into ohmic or Schottky junction. In this work, we consider the general case, that is ohmic at left end and Schottky at right. Due to the depletion region formed at the interface, the current $I_{in}$, therefore, transmits through the device partially, and the transmitted current can be obtained by:

\begin{align}
I_{T}=TI_{in}=-q^2 \frac{V_{e}}{\pi \hbar}T
\label{eq:4}
\end{align}\noindent
where $T$ is the transmission coefficient for an electron that has the Fermi energy in the left electrode. From Eq.~(\ref{eq:4}), it can be seen that the problem of calculating the current in the short device is transformed into a quantum scattering problem. Thus, the key is to analyse the transmission coefficient $T$. 

\begin{figure}
\includegraphics[width=8cm]{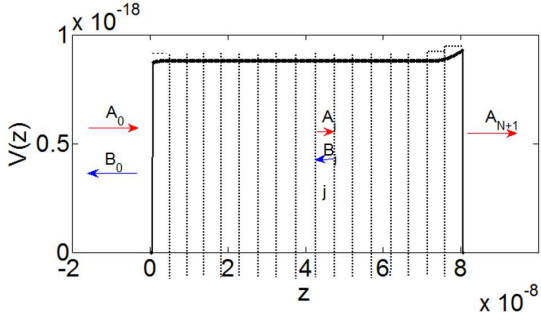}
\caption{\label{fig6} Energy diagram of the device when there is no applied voltage and strain.}
\end{figure}

In order to calculate the $T$, a method that can be used to calculate the transmission coefficient of arbitrary potential barrier is used \cite{ref11}. As shown in the Fig.~\ref{fig6}, the potential inside the wire in $z$-direction has been divided into many small rectangles, the potential $U(z_{j})$ of $j^{th}$ segment is given by $U(z_{j})=V[(z_{j-1}+z_{j})/2]$, where $j$ is the number of the segment($j$=0, 1, 2, 3, … N, N+1), and $V(z)$ is the potential energy of the whole device including the electrodes. If the $j$ is taken larger and larger, the continuous potential variation will be recovered gradually. For a single rectangular potential barrier $j$, the time-independent wavefunction of an electron inside $j$ is described by the Schrodinger equation:

\begin{align}
E\psi(z_{j})+\frac{\hbar^2}{2m} \frac{d^2\psi(z_{j})}{dx^2}-U(z_{j})\psi(z_{j})=0
\label{eq:5}
\end{align}\noindent
where the $E$ is the overall energy of the electron. The wave function of $\psi_{j}$ can be derived easily:

\begin{align}
\psi_{j}=A_{j}e^{ik_{j}z}+B_{j}e^{-ik_{j}z}
\label{eq:6}
\end{align}\noindent
where $k_{j}=\sqrt{2m(E-U(z_{j}))}/\hbar$. According to the quantum theory, the $\psi_{j}$ and $d\psi_{j}/dz$ should be continuous at each boundary, then the amplitude $A_{j}$ and $B_{j}$ can be determined by the following equation \cite{ref11}:

\begin{align}
\left( \begin{array}{c}
A_{j} \\
B_{j}
\end{array} \right)=\prod_{l=0}^{j=1}M_{l}\left(\begin{array}{c}
A_{0} \\
B_{0}
\end{array}\right)
\label{eq:7}
\end{align}\noindent
where $M_{l}$ is given by:

\begin{align}
M_{l}=\left[ \begin{array}{cc}
(1+S_{l})e^{-i(k_{l+1}-k_{l})z_{l}} &(1-S_{l})e^{-i(k_{l+1}+k_{l})z_{l}} \\
(1-S_{l})e^{i(k_{l+1}+k_{l})z_{l}} & (1+S_{l})e^{i(k_{l+1}-k_{l})z_{l}}
\end{array} \right]
\label{eq:8}
\end{align}\noindent
and $S_{l}=k_{l}/k_{l+1}$.The scattering matrix $M$ can then be obtained as:

\begin{align}
M=\left[ \begin{array}{cc}
M_{11} &M_{12} \\
M_{21} &M_{22} 
\end{array} \right]=\prod_{l=0}^{N}M_{l}
\label{eq:9}
\end{align}\noindent
Finally, the $T$ can be calculated by: $T=k_{N+1}/k_{0}|k_{0}/(k_{N+1}M_{22})|^2$, provided the $A_{0}$=1 and $B_{N+1}$=0.

The potential energy function $V(z)$ has to be clarified if calculating the $T$ by the above method. First, the potential variation under an applied $V_{e}$ is assumed to be linear as the quantum wire is not a perfectly metallic wires but a short nanoscale device, and the electrodes have a relatively bigger cross-section area. The source and drain capacitances can then be simply represented by parallel plate capacitors. The potential energy in the wire can be described by:

\begin{align}
V_{w}(z)=V_{w0}-qV_{e}\frac{C_{D}(z)}{C_{ES}(z)}
\label{eq:10}
\end{align}\noindent
where $C_{ES}=C_{D}+C_{S}$ is total electrostatic capacitance at point at $z$, and $C_{D}$ and $C_{S}$ are the capacitance linking the point at $z$ to drain and source, respectively. $V_{w0}$ is the initial potential energy field of the electrons in the quantum wire. We have presented the picture of potential $V_{w0}$ in Fig.~\ref{fig6}, where at the left end and right end the potential variations due to the M-S contact have been considered. The $V_{w0}$ in the M-S contact region can be calculated by Poisson equation, In the simulation, when calculating the potential of the Schottky junction, we use the equation $V_{w0}=-q^2N_{d}/(2\varepsilon_{s})[z_{d}^2-(z_{d}-z)^2]$, where the origin is taken at the right interface, and the direction of $z$ is pointing to the left. As the electron flows from the left to the right, the Schottky at the right end is forwardly biased. The $z_{d}$ is the depletion layer width, and it can be obtained as $z_{d}=\sqrt{2\varepsilon_{s}(\phi_{b}-V_{e})/qN_{d}}$. $N_{d}$ is the donor density and $\phi_{b}$ is the built-in potential which can be calculated easily with the work function of the electrode. Similarly, the variation potential energy in a ohmic junction can be derived by passion equation as well, i.e. $V_{w0}=q^2N_{d}/(2\varepsilon_{s})[z_{0}^2-(z_{0}-z)^2]$, where $z_{0}$ is the width of the charged region where ohmic junction is formed. Apart from the charged region at the two interfaces, the $V_{w0}$ in the rest region can be derived by continuous condition.  

It should be noted that we have ignored the charging effect of the wire in Eq.~\ref{eq:10}, assuming one electron that flows into the nanowire would not affect the local potential. Substituting $C_{S}(z)=\varepsilon A/z$, $C_{D}(z)=\varepsilon A/(L_{z}-z)$  into Eq.~\ref{eq:10}, the $V_{w}(z)$ can then be explicitly expressed as:

\begin{align}
V_{w}(z)&=V_{w0}-qV_{e}\frac{1/C_{S}(z)}{1/C_{D}(z)+1/C_{S}(z)} \nonumber\\
&=V_{w0}-qV_{e}\frac{z}{L_{z}}
\label{eq:11}
\end{align}\noindent
When the substrate of the device is stretched or compressed, there will be strain induced in the quantum wire along the $z$-axis, and due to the piezoelectricity, the quantum wire will generate piezoelectric charge at the interface of the ZnO wire and the electrode. Meanwhile, the induced piezoelectric charges change the local electrical field and the potential. Thus at the left and right interfaces, the piezopotential $V_{p}$ has to be considered according to piezoelectricity theory. Assuming there is a tensile stain $s_{33}$ in z-direction, then negative charge will be induced at the right interface and positive charge will be at the left interface, and if we define the density of the piezoelectric charge induced at the interface is $\rho_{p}$ at the two interfaces, the piezoelectric potential can be obtained by using the Poisson equation:

\begin{align}
-\frac{d^2V_{p}(z)}{dz^2}=q\frac{\rho_{p}(z)}{\varepsilon_{s}}
\label{eq:12}
\end{align}\noindent
Take the left interface for instance, by integrating the Eq.~\ref{eq:12} over the piezoelectric width $z_{pl}$, we can get:

\begin{align}
&V_{p}(z)=\frac{q}{\varepsilon_{s}}\rho_{p}(z_{pl}-z/2)z  \nonumber\\
&z_{l} \leq z \leq z_{pl}
\label{eq:13}
\end{align}\noindent
Overall, under an applied voltage $V_{e}$ , and meanwhile, suffering a strain along the $z$-axis, the potential $V(z)$ in the device can be represented by a piece wise function, that is:

\begin{align}
V(z)=\left\{
  \begin{array}{lr}
    V_{L} &  z < z_{l}\\
    V_{w}+qV_{p}(z) & z_{l} \leq z \leq z_{pl} \\
    V_{w} & z_{pl}<z<z_{pr}\\
    V_{w}+qV_{p}(z) & z_{pr} \leq z \leq z_{N} \\
    V_{R} &  z > z_{N}
  \end{array}
\right.
\label{eq:14}
\end{align}\noindent
where $V_{L}$ and $V_{R}$ are the potential in the two electrodes, $V_{p}(z)$ is the piezoelectric potential generated at the interface, which depends on the strain at the point $z$. $z_{pl}$ and $z_{pr}$ are the assumed width that has piezoelectric charge at the left and right interface, respectively. For the left ($z<z_{0}$) and right ($z>z_{N}$) region, where the potential is constant, the solutions to the Schrodinger equation are plane waves, which are expressed as:

\begin{align}
\psi_{L}(z)=A_{0}e^{ik_{L}z}+B_{0}e^{-ik_{L}z}
\label{eq:15}
\end{align}\noindent
and

\begin{align}
\psi_{R}(z)=A_{N+1}e^{ik_{R}z}+B_{N+1}e^{-ik_{R}z}
\label{eq:16}
\end{align}\noindent
with $k_{L}=\sqrt{2m(E-V_{L})/\hbar}$ and $k_{R}=\sqrt{2m(E-V_{R})/\hbar}$  as the wavenumber, where $E$ is the total energy of the electron. In this work, only the electrons that come from the source are considered so the plane wave that come from drain are ignored, i.e. $B_{N+1}=0$. Combining the Eqs. 6 - 16, the transmission coefficient $T$ ($T=k_{N+1}/k_{0}|k_{0}/(k_{N+1}M_{22})|^2$) of an electron  that has Fermi energy can be numerically calculated.

\section{Numerical Simulation and Discussion}

For the ZnO nanowire, the ballistic channel length $l$ can be evaluated by comparing the transit time $\tau$ and its average scattering time $\tau_{m}$, that is

\begin{align}
\frac{\tau}{\tau_{m}}=\frac{l^2/\mu_{m} V_{e}}{m\mu_{m} /q}=\frac{ql^2}{mV_{e}\mu_{m}^2} 
\label{eq:17}
\end{align}\noindent
where $m$ is the effective mass of the electron and $\mu_{m}$ is electron's mobility. Assuming the $V_{e}$=1 V, $\mu_{m}$=500$cm^2/Vs$ and The $l$ is calculated, when $\tau/\tau_{m}$=1, to be about 92 nm. In our simulation the length of the nanowire $L_{z}$ will be taken as 80nm as to satisfy the ballistic transport regime.

As the $z_{pl}$ is very narrow ($z_{pl} \ll L_{z}$), the variation of the $V_{p}(z)$ can be ignored, and we take $z_{pl}$ equal to the width of one segment. For the right interface, the induced charge is positive so the $V_{p}(z)$ at right interface is negative, and by symmetry we can take the $-V_{pa}$ to represent the piezoelectric voltage over there. As the polarization of the piezoelectric $P_{z}$ can also be expressed as: $P_{z}=e_{33}s_{33}=q\rho_{p}z_{pl}$ , the $V_{pa}$ can then be re-written as: $V_{pa}=z_{pl}e_{33}s_{33}/(2\varepsilon_{s})$, in which the strain and the piezoelectric potential is related \cite{ref8}. 
The strain $s_{33}$ is varying in [-0.5/100, 0.5/100] in this simulation. The piezoelectric constants are taken as $e_{31}$=-0.51 $C/m^2$, $e_{33}$=1.22$C/m^2$ and $e_{15}$=-0.45$C/m^2$, respectively, and the relative dielectric constant $\varepsilon_{s}$=8.91. 
The material of the two electrodes is Ag. The donor density $N_{d}$ in the ZnO is taken as $1\times10^{16}/cm^3$. The built-in potential of the Schottky barrier $\phi_{b}$=0.3 eV. 
An electron that comes from the source with $E=E_{F}\approx 5.49eV$. The potential field in the nanowire is divided into 360 grids ($z_{1}$ to $z_{361}$). 

\begin{figure}
\includegraphics[width=8cm]{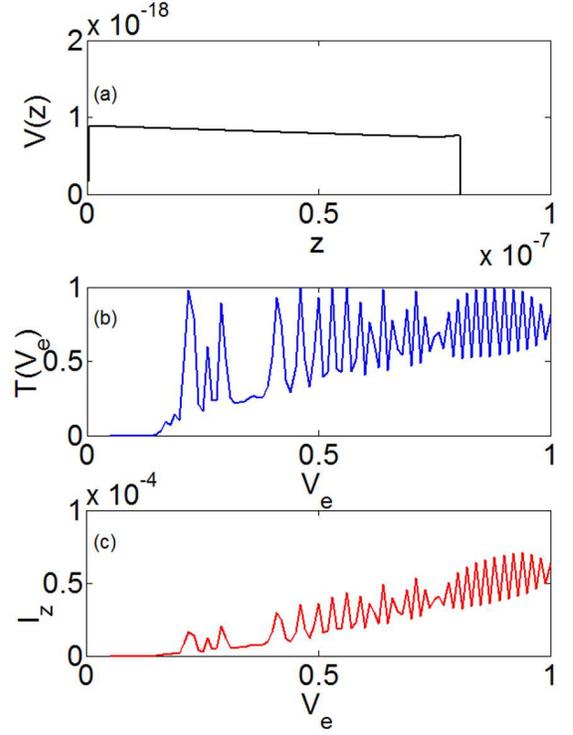}
\caption{\label{fig7} Transmission result when the device is applied a voltage $V_{e}$ (0-1V). a) potential energy. b) Transmission $vs$ $V_{e}$. c) Transmission current $I_{z}$ $vs$ $V_{e}$.}
\end{figure}

\begin{figure}
\includegraphics[width=8cm]{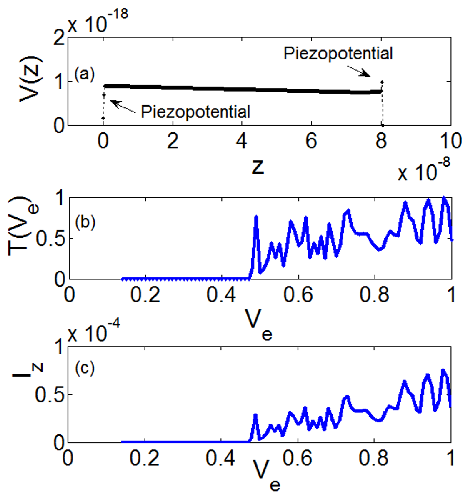}
\caption{\label{fig8} Transmission result when the device is applied a voltage Ve (0-1V) and suffering a stretched strain 05/100. a) potential energy. b) Transmission $vs$ $V_{e}$. c) Transmission current $I_{z}$ $vs$ $V_{e}$.}
\end{figure}

\begin{figure}
\includegraphics[width=8cm]{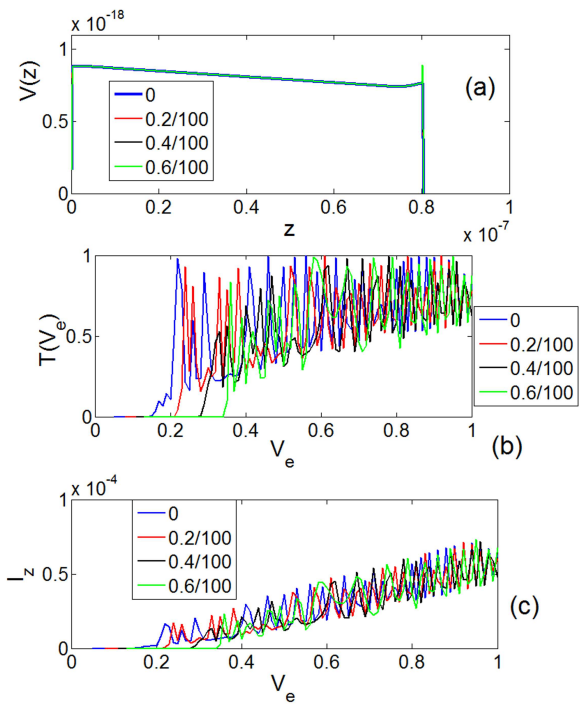}
\caption{\label{fig9} Energy band (a), Transmission $vs$ $V_{e}$ (b), and Transmission current $I_{z}$ $vs$ $V_{e}$ (c) for different strains.}
\end{figure}

\begin{figure}
\includegraphics[width=8cm]{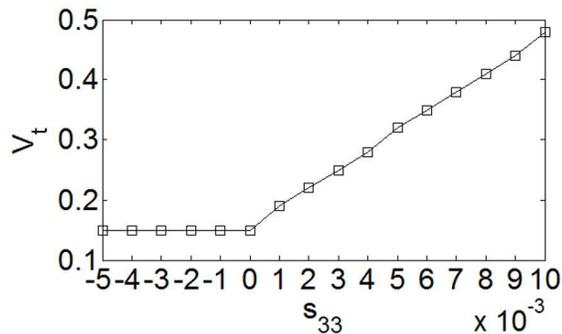}
\caption{\label{fig10} Switch voltage $V_{t}$ when the device suffering a strain from -0.5/100 to 0/5/100.}
\end{figure}

Based on the parameters taken above, the potential energy diagram of the device including two electrodes has been plotted in Fig.~\ref{fig6}, where the Schottky barrier at the right side is about 0.3 eV, and an electron that have energy $E_{F}$ from the left electrode with amplitude $A_{0}$ and $B_{0}$ is shown, represented by the red and blue arrows in the figure. The amplitude of the incoming electron in each segment are presented by $A_{j}$ and $B_{j}$, and when the electron reach the right electrode, it has only the amplitude $A_{N+1}$ as we assume there is no electron coming from right electrode.
The transmission coefficient $T$ when there is no strain in the nanowire is firstly calculated in Fig.~\ref{fig7}. In Fig.~\ref{fig7}(a), the potential energy diagram under the applied voltage $V_{e}$ is shown. The potential is linearly decreased from the left to right, as we added the negative voltage to the left. In Fig.~\ref{fig7}, it is seen the variation of the $T$ versus applied voltage $V_{e}$. The transmission coefficient displays a bit nonlinearity while in some regions the $T$ is oscillating periodically due to the resonance through the virtual states above the barrier. In Fig.~\ref{fig7}(c), the current with the varied applied voltage $V_{e}$ has been plotted by employing the quantum transport Eq.~\ref{eq:4}. Furthermore, when there is a strain of 0.5/100 along the $z$-axis of the device, the potential diagram, transmission coefficient and quantum current have been calculated in Fig.~\ref{fig8}. As we can see in Fig.~\ref{fig8}, due to the strain, the piezopotential is appearing at the two interfaces, indicated by the arrows. Using this method, we can accurately model the piezopotential. In the conventional treatment, the piezopotential can only be assumed in the layer \cite{ref8}. In Figs.~\ref{fig8}(b) and (c) the transmission coefficient $T$ and transmitted current versus $V_{e}$ under the strain of 0.5/100 have been plotted. Compared with Figs.~\ref{fig7}(b) and (c), there is less oscillation, and the switch voltage for opening up the device is increased. That is because the piezopotential induced at the interface have led to the potential inside the device discontinuous as well as making the barrier at the right interface higher. Fig. 9 displays results for energy band, $T$-$V$, and $I$-$V$ for various strains. The results reflect a combination of the asymmetric and symmetric effects. When the strain is 0, there is no piezoelectric effect. Combined piezoelectric and piezoresistance effects appear as the device under a non-zero strain. In order to investigate how the strain affects the switch voltage of the device, we have calculated the switch voltage when the device is suffering the external strain from -0.5/100 to 0.5/100, it is found from Fig.~\ref{fig10} that when the device is compressed the switch voltage keeps constant, while the threshold will be increased linearly when the device is stretched. That is because the current transport in this device is determined by the barrier at the right, and only if the device surfers a stretched strain the barrier at the right interface become higher. Under the quantum regime, the piezopotential is considered as being induced by surface charges and has no significant effect on the Schottky potential in the nanowire body. Therefore as the device under a tensile stress, the piezopotential is larger than the Schottky barrier height (SBH), rising up $V_{t}$. On the contrary, as the piezopotential is smaller than the SBH for the case of the device having a compressive strain, the $V_{t}$ is constant (governed by the SBH).

\section{Conclusion}
As the device dimension reduces to the scale comparable to the de Broglie length, quantum mechanics theory has to be used. In this paper, quantum analysis of the piezotronics of ZnO two-terminal device has been performed. One dimensional chemical potential of the device has been derived including the strain induced charges and the Schottky junction effect. Transmission probability of electrons along the calculated potential has been calculated using the quantum scattering theory. It is found that the threshold point of the gate voltage has been influenced by the piezoelectric effect, which coincides with the results from the conventional theory. Moreover the electrical current fluctuates when the gate voltage is at threshold region due to quantum tunnelling resonance.

\bibliography{refs}

\end{document}